\documentclass[10pt]{amsart}
\usepackage{amssymb, amsmath, amsthm, enumerate, times, hyperref, footnote}

\hypersetup{hyperfootnotes=true}

\begin{document}



\title{A Note on an NSFD Scheme for a Mathematical Model of Respiratory Virus Transmission$^\text{\hyperlink{a}{a}}$\footnote[1]{{$^\text{\lowercase{\hypertarget{a}{a}}}$T\lowercase{he final version will be published in} \textit{J\lowercase{ournal of} D\lowercase{ifference} E\lowercase{quations and} A\lowercase{pplications}}}.\vspace{.1in}}
}

\author{Ronald E. Mickens}
\address{Department of Physics, Clark Atlanta University, Atlanta, GA 30314, USA}
\email{rmickens@cau.edu}

\author{Talitha M. Washington}
\address{Department of Mathematics, University of Evansville, Evansville, IN 47722, USA}
\email{tw65@evansville.edu}


\begin{abstract}
We construct a nonstandard finite difference (NSFD) scheme for an SIRS mathematical model of respiratory virus transmission.  This discretization is in full compliance with the NSFD methodology as formulated by R. E. Mickens.  By use of an exact conservation law satisfied by the SIRS differential equations, we are able to determine the corresponding denominator function for the discrete first-order time derivatives.  Our scheme is dynamically consistent with the SIRS differential equations since the conservation laws are preserved.  Further, the scheme is shown to satisfy a positivity condition for its solutions for all values of the time step-size.
\end{abstract}

\keywords{SIRS models; epidemic models; nonstandard finite difference schemes; nonstandard finite difference schemes; respiratory virus transmission}

\subjclass[2000]{39A50; 65L05; 65L12; 65L20}

\maketitle

\section{Introduction}

In this paper, we will demonstrate how to correctly construct a finite difference scheme for a mathematical model of respiratory virus transmission \cite{WebWebMill} within the general discretization framework of Mickens' nonstandard finite difference (NSFD) methodology \cite{MickNSFD, MickDenom, MickPop}.  Our work clarifies the previous results of A. J. Arenas et al. \cite{AreMorCor}.  While they began their calculations using a version of the NSFD discretization, they did not construct a scheme that is fully consistent with the NSFD methodology.  In particular, they use an ad hoc (but excellent) selection for the denominator functions; further, their scheme places an upper bound on the value of the time step size that can be used for numerical computations.  As we will show, our NSFD scheme has neither of these limitations.

The following system ordinary differential equations have been used to model the transmission of the respiratory synacytical virus (RSV) transmission \cite{AreMorCor, WebWebMill}
\begin{subequations} \label{eq1} \begin{align} \frac{dS}{dt} & = \mu - \mu S(t)-\beta(t)S(t)I(t)+\gamma R(t), \label{eq1a}\\ \frac{dI}{dt} & = \beta(t)S(t)I(t)-\nu I(t)-\mu I(t), \label{eq1b} \\ \frac{dR}{dt} & =\nu I(t)-\gamma R(t)-\mu R(t) \label{eq1c} \end{align}
\noindent where $S(t)$, $I(t)$, and $R(t)$ represent the susceptible, infected and recovered populations, respectively. The initial conditions are
\begin{equation} S(0)=S_0\geq 0, \ I(0)=I_0\geq 0, \ R(0)=R_0\geq 0,\label{1d}\end{equation} \end{subequations}
\noindent and the parameters ($\mu, \nu, \gamma$) are non-negative.  The coefficient function $\beta(t)$ is taken to be periodic and non-negative, and often has the following general form \cite{AreMorCor, WebWebMill}
\begin{subequations}
\begin{equation} {\beta(t)=b_0\left(1+b_1\cos (2\pi t+\phi)\right)} , \label{beta} \end{equation}
\noindent with \begin{equation} 0 \leq b_0, \ 0 \leq b_1\leq 1, \ 0 \leq \phi \leq 1, \end{equation}
\end{subequations}
\noindent where $\phi$ is a normalized (constant) phase angle.
Note that the structure of Equations (\ref{eq1a}), (\ref{eq1b}), and (\ref{eq1c}) is such that \begin{equation}\{S(0)\geq 0, I(0) \geq 0, R(0)\geq 0\} \Rightarrow \{S(t)\geq 0, I(t)\geq 0, R(t)\geq 0\} ,\end{equation} for $t >0$ \cite{Thi} .  We call this the ``positivity condition."
Adding Equations (\ref{eq1a}), (\ref{eq1b}), and (\ref{eq1c}) gives
\begin{subequations}
\begin{equation} \frac{dP}{dt} = \mu(1-P), \ P(0)=S(0)+I(0)+R(0)=P_0 \label{eq4a}\end{equation}
\noindent where \begin{equation} P(t)=S(t)+I(t)+R(t)\label{eq4b} \end{equation}
\end{subequations}
\noindent is the total population.  We call Equation (\ref{eq4a}) the ``conservation law" corresponding to the system of the three ordinary differential equations given in Equation (\ref{eq1}).  The solution to Equation (\ref{eq4a}) is \cite{MickPop} \begin{equation} P(t)=1+(P_0-1)e^{-\mu t} .\end{equation}  This result, along with the positivity requirement, allows us to conclude that if $0 \leq S_0\leq 1, 0 \leq I_0 \leq 1, 0 \leq R_0 \leq 1$, with $0 \leq P_0\leq 1$, then \begin{equation}0 \leq S(t)\leq 1, \ 0\leq I(t)\leq 1, \ 0 \leq R(t)\leq 1 .\end{equation}  Thus, we not only have positivity, but also boundedness of the solutions.  (Note that $P(t)$ is not necessarily a constant function which is what is stated in A. J. Arenas et al. \cite{AreMorCor}.)

The differential equations (\ref{eq1a}), (\ref{eq1b}), (\ref{eq1c}) and (\ref{eq4a}) constitute a member of the class of population models satisfy a conservation law.  Utilizing the procedures for constructing the appropriate NSFD scheme for such systems as shown in \cite{MickPop} gives \begin{equation*} \begin{aligned} &t \rightarrow t_N=hN , \ h=\Delta t;\\ & \left(S(t), I(t), R(t)\right) \rightarrow \left(S_N, I_N, R_N\right), \\ &\beta(t) \rightarrow \beta(t_N)=\beta_N . \end{aligned} \end{equation*}
which is what is not implemented in the scheme shown in A. J. Arenas et al. \cite{AreMorCor}.

First, we observe that Equation (\ref{eq4a}) has the following exact finite difference scheme \cite{MickNSFDMod, MickNSFD, MickPop} \begin{equation} \frac{P_{N+1}-P_N}{\phi}=\mu(1-P_{N+1}) , \end{equation} where the denominator function \cite{MickNSFDMod} is \begin{equation} \phi(\mu,h)=\frac{e^{\mu h}-1}{\mu} . \label{eq8} \end{equation}  As explained in \cite{MickPop}, this denominator function must be used in the discretization of Equations (\ref{eq1a}), (\ref{eq1b}), and (\ref{eq1c}), otherwise, the exact NSFD scheme for the total population conservation law will not hold.  In a similar manner, again see Mickens \cite{MickPop} for the details, the NSFD scheme for the three original differential equations, must take the form
\begin{subequations} \label{eq9}
\begin{align} \frac{S_{N+1}-S_N}{\phi}&=\mu-\mu S_{N+1}-\beta_NS_{N+1}I_N+\gamma R_{N+1}, \label{eq9a} \\
\frac{I_{N+1}-I_N}{\phi}&=\beta_NS_{N+1}I_N-\nu I_{N+1}-\mu I_{N+1} , \label{eq9b} \\
\frac{R_{N+1}-R_N}{\phi}&=\nu I_{N+1}-\gamma R_{N+1}-\mu R_{N+1} . \label{eq9c} \end{align}
\end{subequations}
\noindent Defining $P_N \equiv S_N+I_N+R_N$ and adding these equations gives \begin{equation*} \frac{P_{N+1}-P_N}{\phi}=\mu(1-P_{N+1}) , \ P_0=S_0+I_0+R_0 , \end{equation*} which is the exact finite scheme for the conservation law as expressed by Equation (\ref{eq4a}).  Note that the solution for $P_N$ is \cite{SpiMoy} \begin{equation} P_N=1+(P_0-1)e^{-\mu h N} .\end{equation}  Thus, we can immediately reach the following conclusion:
\begin{enumerate}
\item[]
If, $0\leq S_0 \leq 1 , 0 \leq I_0 \leq 1,$ and $0 \leq R_0 \leq 1 $ with $ {0} \leq P_0 \leq 1 ,$

\noindent then $0 \leq P_N \leq 1 , 0 \leq S_N \leq 1 , 0 \leq I_N \leq 1, $ and $0 \leq R_N \leq 1$ for all $N \geq 1$ .
\end{enumerate}
\noindent Consequently, this NSFD scheme gives numerical solutions that have the correct upper bound and they satisfying the positivity condition.

The system of difference equations given in Equations (\ref{eq9}) are not in a form suitable for computation. Since each of these equations is linear in $S_{N+1}, I_{N+1}, $ and $R_{N+1}$, a rather long but elementary calculation gives \cite{SpiMoy} \begin{equation} S_{N+1} = \frac{B_N}{A_N} , \quad I_{N+1}=\frac{C_N}{A_N} , \quad R_{N+1}=\frac{D_N}{A_N} , \end{equation} where \vspace{-.05in}
\begin{subequations} \label{eq12}
\begin{align} A_N &= e^{3\mu h}+\phi[(\nu+\gamma)+\beta_N I_N]e^{2\mu h}+\phi^2[\gamma \nu+(\nu + \gamma)\beta_N I_N]e^{\mu h} , \label{eq12a} \\[7pt]
B_N & = (\mu \phi + S_N)[(1+\mu \phi)+\nu \phi][(1+\mu \phi)+\gamma \phi]\cr
 & +(\gamma \nu \phi^2)I_N+(\gamma \phi)[(1+\mu \phi)+\nu \phi]R_N \label{eq12b} \\[7pt]
C_N & = \{[(1+\mu \phi)+\phi \beta_N {I_N}][(1+\mu \phi)+\gamma \phi]+(\phi^2\gamma \beta_N)R_N \cr
&+ (\phi \beta_N)(\mu \phi+S_N)[(1+\mu\phi)+\gamma \phi]\}I_N , \label{eq12c} \\[7pt]
D_N & = [(1+\mu\phi)+\phi \beta_N I_N][(1+\mu \phi)+\nu \phi]R_N \cr
&+(\phi^2\nu\beta_N)(\mu\phi+S_N)I_N + (\nu \phi)[(1+\mu\phi)+\phi\beta_NI_N]I_N , \label{eq12d} \end{align}
\end{subequations}
\noindent and $\phi=\phi(\mu,h)$ is the denominator function as expressed in Equation (\ref{eq8}).  Inspection of Equations (\ref{eq12}) shows that \begin{equation} A_N > 0 , \quad B_N > 0 , \quad C_N > 0 , \quad \rm{ and } \quad D_N > 0  \label{eq13} \end{equation} for all $N \geq 0$, and for all step-sizes, $h>0$.  Thus, the above constructed NSFD scheme places no limitation on the step-size.


We have achieved the stated purpose of this paper, namely, the correct construction of a NSFD scheme for an SIRS model of RSV.  Our results demonstrate the power of the general NSFD methodology \cite{MickNSFDMod} when applied to systems satisfying a condition of positivity and also having a conservation law \cite{MickPop}.  While this NSFD scheme may not be difficult to formulate, a successful implementation of a NSFD scheme requires that the construction of the scheme agrees with NSFD methodology.  Our NSFD scheme has the following significant features:
\begin{enumerate}
\item The values of $(S_{N+1}, I_{N+1}, R_{N+1} )$ are determined only by $(S_N, I_N, R_N)$, the step size $h$, and the non-negative parameters $(\mu, \gamma, \nu, b_0, b_1)$.

\item The denominator function is explicitly determined and does not require a particular specialized form.

\item The scheme's solution satisfies the positivity requirement, the appropriate boundedness conditions, and maintains these properties for all $h>0$.
\end{enumerate}
Finally, if we take as given the validity of the above stated SIRS mathematical model for RSV transmission \cite{AreMorCor, WebWebMill}, then the NSFD discretization of this paper can be used to study in detail the dynamics of this epidemiological system.


\label{lastpage}

\end{document}